\documentclass[twocolumn,amsmath,showpacs,superscriptaddress]{revtex4}

\usepackage{amsfonts}
\usepackage{amssymb}
\usepackage{graphicx}

\begin{document}

%%%%%%%%%%%%%%%%%%%%%%%%%%%%%%%%%%%%%%%%%%%%%%%%%%%%%%%%%%%%%%%%%%
\title{Signum Function Method for Generation of Correlated Dichotomic Chains}
%%%%%%%%%%%%%%%%%%%%%%%%%%%%%%%%%%%%%%%%%%%%%%%%%%%%%%%%%%%%%%%%%%

%%%%%%%%%%%%%%%%%%%%%%%%%%%%%%%%%%%%%%%%%%%%%%%%%%%%%%%%%%%%%%%%%%
\author{S.~S.~Apostolov}

\affiliation{A.~Ya.~Usikov Institute for Radiophysics and
             Electronics, Ukrainian Academy of Science,
             12 Proskura Street, 61085 Kharkov, Ukraine}
%%%%%%%%%%%%%%%%%%%%%%%%%%%%%%%%%%%%%%%%%%%%%%%%%%%%%%%%%%%%%%%%%%

%%%%%%%%%%%%%%%%%%%%%%%%%%%%%%%%%%%%%%%%%%%%%%%%%%%%%%%%%%%%%%%%%%
\author{F.~M.~Izrailev}
\email{izrailev@venus.ifuap.buap.mx}

\affiliation{Instituto de F\'{\i}sica, Universidad Aut\'{o}noma de Puebla,\\
             Apartado Postal J-48, Puebla, Pue., 72570, M\'{e}xico}
%%%%%%%%%%%%%%%%%%%%%%%%%%%%%%%%%%%%%%%%%%%%%%%%%%%%%%%%%%%%%%%%%%

%%%%%%%%%%%%%%%%%%%%%%%%%%%%%%%%%%%%%%%%%%%%%%%%%%%%%%%%%%%%%%%%%%
\author{N.~M.~Makarov}
\email{makarov@siu.buap.mx}
\thanks{On sabbatical leave from Instituto de Ciencias,
        Universidad Aut\'{o}noma de Puebla,
        Priv. 17 Norte No. 3417, Col. San Miguel
        Hueyotlipan, Puebla, Pue., 72050, M\'{e}xico.}

\affiliation{A.~Ya.~Usikov Institute for Radiophysics and
             Electronics, Ukrainian Academy of Science,
             12 Proskura Street, 61085 Kharkov, Ukraine}
%%%%%%%%%%%%%%%%%%%%%%%%%%%%%%%%%%%%%%%%%%%%%%%%%%%%%%%%%%%%%%%%%%

%%%%%%%%%%%%%%%%%%%%%%%%%%%%%%%%%%%%%%%%%%%%%%%%%%%%%%%%%%%%%%%%%%
\author{Z.~A.~Mayzelis}

\affiliation{A.~Ya.~Usikov Institute for Radiophysics and
             Electronics, Ukrainian Academy of Science,
             12 Proskura Street, 61085 Kharkov, Ukraine}
%%%%%%%%%%%%%%%%%%%%%%%%%%%%%%%%%%%%%%%%%%%%%%%%%%%%%%%%%%%%%%%%%%

%%%%%%%%%%%%%%%%%%%%%%%%%%%%%%%%%%%%%%%%%%%%%%%%%%%%%%%%%%%%%%%%%%
\author{S.~S.~Melnyk}

\affiliation{A.~Ya.~Usikov Institute for Radiophysics and
             Electronics, Ukrainian Academy of Science,
             12 Proskura Street, 61085 Kharkov, Ukraine}
%%%%%%%%%%%%%%%%%%%%%%%%%%%%%%%%%%%%%%%%%%%%%%%%%%%%%%%%%%%%%%%%%%

%%%%%%%%%%%%%%%%%%%%%%%%%%%%%%%%%%%%%%%%%%%%%%%%%%%%%%%%%%%%%%%%%%
\author{O.~V.~Usatenko}
\email{usatenko@ire.kharkov.ua}

\affiliation{A.~Ya.~Usikov Institute for Radiophysics and
             Electronics, Ukrainian Academy of Science,
             12 Proskura Street, 61085 Kharkov, Ukraine}
%%%%%%%%%%%%%%%%%%%%%%%%%%%%%%%%%%%%%%%%%%%%%%%%%%%%%%%%%%%%%%%%%%

\date{\today}

%%%%%%%%%%%%%%%%%%%%%%%%%%%%%%%%%%%%%%%%%%%%%%%%%%%%%%%%%%%%%%%%%%
\begin{abstract}
We analyze the signum-generation method for creating random
dichotomic sequences with prescribed correlation properties. The
method is based on a binary mapping of the convolution of continuous
random numbers with some function originated from the Fourier
transform of a binary correlator. The goal of our study is to reveal
conditions under which one can construct binary sequences with a
given pair correlator. Our results can be used in the construction
of superlattices and waveguides with selective transport properties.
\end{abstract}
%%%%%%%%%%%%%%%%%%%%%%%%%%%%%%%%%%%%%%%%%%%%%%%%%%%%%%%%%%%%%%%%%%


%\preprint{Preprint: Sing-Sin}

\maketitle

%%%%%%%%%%%%%%%%%%%%%%%%%%%%%%%%%%%%%%%%%%%%%%%%%%%%%%%%%%%%%%%%%%
\section{Introduction}
\label{sec-Intro}
%%%%%%%%%%%%%%%%%%%%%%%%%%%%%%%%%%%%%%%%%%%%%%%%%%%%%%%%%%%%%%%%%%

The study of properties of disordered complex systems with spatial
and/or temporal correlations is one of the hot topics in modern
physics. Recently, much attention was paid to the related problem of
how to construct disordered materials with specific transport
properties that are due to underlying correlations in a disorder.
One of the important applications of this problem is a creation of
electron nano-devices, optic fibers, rough surfaces, acoustic and
electromagnetic waveguides with selective transport properties.

It is known (see, for instance,
\cite{LGPb88,MakYur89FMYu90,GrinFish89,MakL99}) that many of
properties of systems with weak disorder are determined by the
binary (pair or two-point) correlation function of a corresponding
random process. Recently it was found \cite{ML98,IzKr99,IzMak05}
that specific long-range correlations in disordered potentials can
lead to anomalous transport properties. To date, there exist a
number of algorithms for generating long-range correlated sequences
with prescribed correlations. Among such algorithms the most
widespread one is the \emph{convolution
method}~\cite{Rice4454,Saupe88Feder88,
Peng91,Prak92,Czir95,WOD95,IzKr99,IzMak0103,IzMak04,IzMak05,IKMU07}.
In this method random elements in the generated chain can be of any
value from $-\infty$ to $\infty$. However, in many applications it
is more convenient to construct the sequences with finite number of
random values. An important example of such system, being occurred
in nature, is a sequence of nucleotides in a DNA molecule,
consisting of four elements only.

The simplest case of a random sequence of finite elements is a
stochastic \emph{dichotomic} (binary) chain of only two elements. In
contrast with the case of sequences of continuous elements, the
problem of a construction of binary sequences with given correlation
properties turns out to be tricky. As was recently shown in
Ref.\cite{IKMU07}, there is a serious restriction on the type of
pair correlators in binary sequences, in contrast with the sequences
with continuous distribution of their elements. Many of related
results for binary sequences can be found in
Refs.\cite{MUYG05,MUY06,Yo07}.

A direct way to create dichotomic correlated sequences is to apply
the signum function to the sequence of continuous values, obtained
with the convolution method. This method is based on the convolution
of a white-noise sequence with some function that is determined by
the desired pair correlator. However, as was numerically found in
\cite{CarpBerIvStanley-Nat02}, the created binary sequence turns out
to have the pair correlator different from the expected one. To
date, it remains unclear how to construct binary sequences having
the same pair correlators as in the sequences with continuous values
of their elements. In spite of its quite simple form, the signum
function method is not rigorously analyzed in the literature.

The present paper makes an effort to clarify this problem. Our aim
is to understand under what conditions the discussed method allows
to construct binary sequences with a desired pair correlator. We
perform a detailed theoretical analysis of the restrictions arising
for binary sequences with long-range correlations. Main attention is
paid to the step-wise power spectrum resulting in a power decay of
correlations. This type of correlations is extremely important in
various applications, such as a creation of devices with selective
transport properties.

The paper is organized as follows. Section 2 is devoted to general
properties of binary sequences. In particular, we derive some of the
conditions restricting the form of pair correlators that a binary
sequence can have. In Section 3 we describe in details the
signum-generation method and derive basic relations needed for a
further analysis. In next Section 4 we analyze the case of a
balanced (unbiased) dichotomic sequence, i.e., the sequence with the
zero mean-value. Here we also display the restrictions related to
the discussed method. Our main findings are reported in Section 5
where we consider the most interesting case of long-range
correlations with the step-wise spectrum. In Section 6 we ask a
question about general restrictions appearing in the case of power
decay of correlations for binary sequences. In last Section we give
some additional remarks concerning binary sequences, and summarize
our results. The Appendices contain some of details of analytical
and numerical calculations.

%%%%%%%%%%%%%%%%%%%%%%%%%%%%%%%%%%%%%%%%%%%%%%%%%%%%%%%%%%%%%%%%%%
\section{Necessary Conditions}
\label{sec-NecCond}
%%%%%%%%%%%%%%%%%%%%%%%%%%%%%%%%%%%%%%%%%%%%%%%%%%%%%%%%%%%%%%%%%%

Let us start with generic properties of dichotomic sequences, not
associated with specific choice of a generation method. In view of
constructing the sequences with a given pair correlator, it is of
great importance to know what are restrictions on the type of
correlators allowed for dichotomic sequences. To shed light on this
problem, here we consider a statistically homogeneous random
sequence of symbols $s_n$ consisting of the values $``-1"$ and
$``1"$,
\begin{equation}\label{BinSeqS}
s_n=\{-1,1\},\qquad n\in\textbf{\textbf{Z}}
=\ldots,-2,-1,0,1,2,\ldots
\end{equation}
The canonical definition of the correlation function reads as
follows
\begin{equation}\label{def_cor}
C_s(r)\equiv\overline{s_{n}s_{n+r}}-\overline{s}^2= C_s(0)K_s(r),
\end{equation}
where $\overline{s}\equiv\overline{s_n}$ and
$C_s(0)\equiv\overline{s^2_n}-\overline{s}^2$ are the mean value and
variance of $s_n$, respectively. Taking into account a peculiar
property
\begin{equation}\label{s2=1}
\overline{s_n^2}=s_n^2=1
\end{equation}
of our dichotomic sequence $s_n$, one obtains
\begin{equation}\label{Var-s}
C_s(0)=1-\overline{s}^{\,2}.
\end{equation}
It should be emphasized that the direct relation between the mean
value and higher one-point moments is a specific property of {\it
any} dichotomic chain. For instance, for the binary chain
$\varepsilon_n$ consisting of $``0"$ and $``1"$ we have the relation
$C_\varepsilon(0)=\overline{\varepsilon}(1-\overline{\varepsilon})$
since $\varepsilon_n^2=\varepsilon_n$ in this case. On the contrary,
for a sequence of continuous random numbers of the Gaussian type the
variance and mean value are independent parameters .

To proceed with the two-point moments, we associate the correlator
$C_s(r)$ with the probabilities of two symbols with the same or
opposite signs, occurring at the distance $r$,
\begin{subequations}\label{Pxy}
\begin{eqnarray}
4P(\pm1,\underbrace{\ldots}_{r-1},\pm1)&=&
(\overline{s}\pm1)^2+C_s(r),\label{P11}\\[6pt]
4P(\pm1,\underbrace{\ldots}_{r-1},\mp1)&=&1-\overline{s}^2-C_s(r).
\label{P10}
\end{eqnarray}
\end{subequations}
These relationships also are peculiar properties solely of
dichotomic sequences, they are drastically distinct from the
analogous relations for other random processes (see, e.g.,
Eq.~\eqref{Gauss2PD} for the two-point probability density of the
Gaussian process). This fact is strictly confirmed by
straightforward calculations of Eqs.~\eqref{Pxy}.

In order to obtain the expression \eqref{P11}, one should write the
average $\overline{(s_{n}\pm1)(s_{n+r}\pm1)}$ via the correlator
\eqref{def_cor},
\begin{eqnarray}
\overline{(s_{n}\pm1)(s_{n+r}\pm1)}&=&\overline{s_ns_{n+r}}\pm\overline{s_n}
\pm\overline{s_{n+r}}+1\nonumber\\[6pt]
&=&C_s(r)+(\overline{s}\pm1)^2.\label{ss11}
\end{eqnarray}
On the other hand, the same average can be calculated with the use
of two-symbol probabilities,
\begin{eqnarray}
\overline{(s_n+1)(s_{n+r}+1)}=2\cdot2\cdot
P(1,\underbrace{\ldots}_{r-1},1)\nonumber\\[6pt]
+0\cdot2\cdot P(-1,\underbrace{\ldots}_{r-1},1)+2\cdot0\cdot
P(1,\underbrace{\ldots}_{r-1},-1)\nonumber\\[6pt]
+0\cdot0\cdot P(-1,\underbrace{\ldots}_{r-1},-1)=4
P(1,\underbrace{\ldots}_{r-1},1).\label{ss12}
\end{eqnarray}
\begin{eqnarray}
\overline{(s_n-1)(s_{n+r}-1)}=0\cdot0\cdot
P(1,\underbrace{\ldots}_{r-1},1)\nonumber\\[6pt]
+(-2)\cdot0\cdot P(-1,\underbrace{\ldots}_{r-1},1)+0\cdot(-2)\cdot
P(1,\underbrace{\ldots}_{r-1},-1)\nonumber\\[6pt]
+(-2)\cdot(-2)\cdot P(-1,\underbrace{\ldots}_{r-1},-1)=4
P(-1,\underbrace{\ldots}_{r-1},-1).\label{ss13}
\end{eqnarray}
Then, the combination of Eqs.~\eqref{ss11}, \eqref{ss12}, and
\eqref{ss13} results in the equality \eqref{P11}.

Similarly, in order to derive the expression \eqref{P10}, one can
write,
\begin{eqnarray}
\overline{(s_{n}\pm1)(s_{n+r}\mp1)}&=&\overline{s_ns_{n+r}}
\mp\overline{s_n}\pm\overline{s_{n+r}}-1\nonumber\\[6pt]
&=&C_s(r)+\overline{s}^2-1.\label{ss21}
\end{eqnarray}
Again, the above average can be calculated employing two-symbol
probabilities. Specifically,
\begin{eqnarray}
\overline{(s_n+1)(s_{n+r}-1)}=2\cdot0\cdot
P(1,\underbrace{\ldots}_{r-1},1)\nonumber\\[6pt]
+0\cdot0\cdot P(-1,\underbrace{\ldots}_{r-1},1)+2\cdot(-2)\cdot
P(1,\underbrace{\ldots}_{r-1},-1)\nonumber\\[6pt]
+0\cdot(-2)\cdot P(-1,\underbrace{\ldots}_{r-1},-1)=-4
P(1,\underbrace{\ldots}_{r-1},-1).\label{ss22}
\end{eqnarray}
\begin{eqnarray}
\overline{(s_n-1)(s_{n+r}+1)}=0\cdot2\cdot
P(1,\underbrace{\ldots}_{r-1},1)\nonumber\\[6pt]
+(-2)\cdot2\cdot P(-1,\underbrace{\ldots}_{r-1},1)+0\cdot0\cdot
P(1,\underbrace{\ldots}_{r-1},-1)\nonumber\\[6pt]
+(-2)\cdot0\cdot P(-1,\underbrace{\ldots}_{r-1},-1)=-4
P(-1,\underbrace{\ldots}_{r-1},1).\label{ss23}
\end{eqnarray}
From Eqs.~\eqref{ss21}, \eqref{ss22}, and \eqref{ss23} it follows
the equality \eqref{P10}.

Now, with the use of expressions \eqref{P11} we express the
correlation function $C_s(r)$ via the probabilities to occur three
symbols,
\begin{equation}%\label{}
C_s(r)+(\overline{s}\pm1)^2=4\sum_{a=-1,1}
P(\pm1,\underbrace{\ldots}_{r'-1},a,\underbrace{\ldots}_{r-r'-1},\pm1).
\end{equation}
Probability
$P(\pm1,\underbrace{\ldots}_{r'-1},a,\underbrace{\ldots}_{r-r'-1},\pm1)$
is smaller than both probabilities
$P(\pm1,\underbrace{\ldots}_{r'-1},a)$ and
$P(a,\underbrace{\ldots}_{r-r'-1},\pm1)$. Thus, we can write,
\begin{eqnarray}%\label{}
&&C_s(r)+(\overline{s}\pm1)^2\nonumber\\[6pt]
&&\leqslant4\sum_{a=-1,1}\min\big\{P(\pm1,\underbrace{\ldots}_{r'-1},a),
P(a,\underbrace{\ldots}_{r-r'-1},\pm1)\big\}\nonumber\\[6pt]
&&=\min\big\{1-\overline{s}^2-C_s(r'),1-\overline{s}^2-C_s(r-r')\big\}
\nonumber\\[6pt]
&&+\min\big\{(\overline{s}\pm1)^2+C_s(r'),(\overline{s}\pm1)^2+C_s(r-r')\big\}
\nonumber\\[6pt]
&&=1-\overline{s}^2+\min\big\{-C_s(r'),-C_s(r-r')\big\}\nonumber\\[6pt]
&&+(\overline{s}\pm1)^2+\min\big\{C_s(r'),C_s(r-r')\big\}.
\end{eqnarray}
Here we again have used Eqs.~\eqref{Pxy}. Then, according to the
evident relation
\begin{equation}\label{}
\min\{x,y\}+\min\{-x,-y\}=-|x-y|,
\end{equation}
we arrive at the condition
\begin{equation}\label{ineq1}
|C_s(r')-C_s(r-r')|+C_s(r)\leqslant 1-\overline{s}^2.
\end{equation}
Finally, it is convenient to rewrite Eq.~\eqref{ineq1} for the
normalized correlator $K_s(r)$,
\begin{eqnarray}\label{norm_in1}
|K_s(r')-K_s(r-r')|+ K_s(r)\leqslant1.
\end{eqnarray}
Although we have derived this inequality for $0<r'<r$, a simple
analysis reveals its validity for arbitrary values of $r$ and $r'$.
We would like to stress that the condition \eqref{norm_in1} is
applicable for any mean value $\overline{s}$ of the dichotomic
sequence $s_n$. However, without a loss of generality in what
follows we consider binary sequences with the zero mean, since the
statistical properties of considered sequences do not depend on mean
values.

In a similar manner one can obtain second inequality with the use of
Eq.~\eqref{P10},
\begin{eqnarray}
&&1-\overline{s}^2-C_s(r)=4\sum_{a=-1,1}
P(\pm1,\underbrace{\ldots}_{r'-1},a,\underbrace{\ldots}_{r-r'-1},\mp1)
\nonumber\\[6pt]
&&\leqslant
4\sum_{a=-1,1}\min\big\{P(\pm1\underbrace{\ldots}_{r'-1},a),
P(a,\underbrace{\ldots}_{r-r'-1},\mp1)\big\}\nonumber\\[6pt]
&&=\min\big\{1-\overline{s}^2-C_s(r'),(\overline{s}\mp1)^2+C_s(r-r')\big\}
\nonumber\\[6pt]
&&+\min\big\{(\overline{s}\pm1)^2+C_s(r'),1-\overline{s}^2-C_s(r-r')\big\}.
\label{ineq2}
\end{eqnarray}
Since $C_s(r)=K_s(r)$ in the case of $\overline{s}=0$,
Eq.~\eqref{ineq2} gets a simpler form,
\begin{equation}\label{norm_in2}
|K_s(r')+K_s(r-r')|-K_s(r)\leqslant1\quad\mbox{for}\quad
\overline{s}=0.
\end{equation}

Thus, applying the necessary conditions \eqref{norm_in1} and
\eqref{ineq2}, or \eqref{norm_in2} if $\overline{s}$=0, one can
identify the functions that {\it can not be} treated as binary
correlators of a dichotomic sequence.

Inequalities \eqref{norm_in1}, \eqref{ineq2}  and \eqref{norm_in2}
have to be met for any values of indices $r$ and $r'$ and,
therefore, they actually represent an infinite set of necessary
conditions. Evidently, in every particular case one should choose
the strongest condition. On the other hand, Eqs.~\eqref{norm_in1},
\eqref{ineq2} and \eqref{norm_in2} are automatically satisfied if
one of the indices equals zero, $r=0$ or $r'=0$. The same takes
place when $r=r'$. Summarizing all these facts, we can combine
Eqs.~\eqref{norm_in1} and \eqref{norm_in2} as follows
\begin{eqnarray}\label{normNecCond}
&&\max\left\{|K_s(r')\pm K_s(r-r')|\mp K_s(r)\right\}\leqslant1,
\\[6pt]
&&r\neq0,\quad r'\neq0,\quad r\neq r'
\qquad\mbox{for}\quad\overline{s}=0.\nonumber
\end{eqnarray}
The symbol $\max\{\ldots\}$ implies the absolute maximum of a
function with respect to the indices $r$, $r'$. Note that
Eq.~\eqref{normNecCond} is automatically fulfilled for two limit
cases, namely, for the delta-correlated (white noise) chain, and for
the sequences with infinitely long-range correlations. As one can
see, for both these cases $K_s(r)=\delta_{r,0}$ or $K_s(r)=1$,
respectively.

%%%%%%%%%%%%%%%%%%%%%%%%%%%%%%%%%%%%%%%%%%%%%%%%%%%%%%%%%%%%%%%%%%
\section{Binary versus Gaussian}
\label{sec-BinGauss}
%%%%%%%%%%%%%%%%%%%%%%%%%%%%%%%%%%%%%%%%%%%%%%%%%%%%%%%%%%%%%%%%%%

Now we analyze the construction of a dichotomic sequence $\gamma_n$
by means of the {\it signum function generation} (SFG) method. It
uses an intermediate correlated disorder $\beta_{n}$ obtained as a
convolution of the uncorrelated Gaussian noise $\alpha_{n}$ and
modulation function $G(n)$. Specifically, the $\gamma$-sequence is
defined by
\begin{subequations}\label{gamma-beta}
\begin{eqnarray}
\gamma_n&=&\mathrm{sign}(\beta_{n}),
\label{gamma}\\[6pt]
\beta_{n}&=&\overline{\beta}+\sum_{n'=-\infty}^\infty
G(n-n')\alpha_{n'}.\label{beta}
\end{eqnarray}
\end{subequations}

The initial Gaussian white-noise chain consists of stochastic
variables $\alpha_n$ with the zero mean, unit variance and
corresponding probability density. Respectively,
\begin{subequations}\label{alpha}
\begin{eqnarray}
&&\overline{\alpha}=0,\qquad\qquad
\overline{\alpha_n\alpha_{n'}}=\delta_{n,n'}\,,\\[6pt]
&&\rho_A(\alpha_n=\alpha)=\frac{1}{\sqrt{2\pi}}\exp(-\alpha^2/2).
\label{rho-alpha}
\end{eqnarray}
\end{subequations}
The bar over a random symbol or function implies the stochastic
average.

Evidently, the constructed dichotomic sequence $\gamma_n$ does not
change if one normalizes the intermediate sequence $\beta_{n}$ by an
arbitrary factor. Hence, without any loss of generality, we can
admit the variance of $\beta_{n}$ be equal to unity. Then,
$\overline{\beta}$ is the mean value of $\beta_{n}$ and its
correlation function $K_{\beta}(r)$,
\begin{equation}\label{Kbeta}
K_{\beta}(r)=\overline{(\beta_{n}-\overline{\beta}\,)
(\beta_{n+r}-\overline{\beta}\,)},
\end{equation}
is normalized to unity, $K_{\beta}(0)=1$. By direct substitution of
Eq.~\eqref{beta} into definition \eqref{Kbeta}, the correlator
$K_{\beta}(r)$ is readily associated with the modulation function
$G(n)$,
\begin{equation}\label{KG-rel}
K_\beta(r)=\sum_{n=-\infty}^\infty G(r-n)G(n).
\end{equation}
Owing to evenness of $K_\beta(r)=K_\beta(-r)$ and in accordance with
Eq.~\eqref{KG-rel}, one can also restrict the function $G(n)$ to the
class of even functions, $G(-n)=G(n)$. Note that the condition
$K_\beta(0)=1$ gives rise to the following normalization for $G(n)$,
\begin{equation}\label{G-Norm}
K_{\beta}(0)=\sum_{n=-\infty}^{\infty}G^2(n)=1.
\end{equation}

It is convenient to pass to the Fourier transform of
Eq.~\eqref{KG-rel} with the use of the standard expressions,
\begin{subequations}\label{FTr-K}
\begin{eqnarray}
&&K_\beta(r)=\frac{1}{2\pi}\int_{-\pi}^{\pi}dk\,
\mathcal{K}_\beta(k)\exp(ikr),\label{FTr-Kr}\\[6pt]
&&\mathcal{K}_\beta(k)=\sum_{r=-\infty}^{\infty}K_\beta(r)\exp(-ikr).
\label{FTr-Kk}
\end{eqnarray}
\end{subequations}
The Fourier transform $\mathcal{K}_\beta(k)$ of the pair correlator
$K_\beta(r)$ introduced here, is known as the \emph{power spectrum}
of random $\beta$-chain. Since the correlator $K_\beta(r)$ is real
and even function of $r$, the power spectrum \eqref{FTr-Kk} is real,
even, $\mathcal{K}_\beta(-k)=\mathcal{K}_\beta(k)$, and non-negative
function of the wave number $k$. Analogously to Eq.~\eqref{FTr-K},
for the modulation function $G(n)$ one can define its Fourier
transform $\mathcal{G}(k)$, which is also real and even function,
$\mathcal{G}(-k)=\mathcal{G}(k)$.

The Fourier representation of Eq.~\eqref{KG-rel} reads
\begin{equation}\label{FTr-KG}
\mathcal{K}_{\beta}(k)=\mathcal{G}^2(k).
\end{equation}
Thus, we arrive at the following expression for the modulation
function $G(n)$,
\begin{equation}\label{G}
G(n)=\frac{1}{\pi}\int_{0}^{\pi}dk\,
\mathcal{K}_\beta^{1/2}(k)\cos(kn).
\end{equation}
Evidently, the solution (\ref{G}) automatically satisfies the
normalization condition (\ref{G-Norm}).

Since the initial chain $\alpha_n$ is a delta-correlated Gaussian
noise, the intermediate variables $\beta_n$ also constitute a
Gaussian random sequence with single-point distribution function
$\rho_B(\beta)$ (see Appendix~\ref{app-RhoBeta}),
\begin{equation}\label{rho-beta}
\rho_B(\beta_n=\beta)=\frac{1}{\sqrt{2\pi}}
\exp\Big[-\big(\beta-\overline{\beta}\big)^2/2\Big].
\end{equation}

In order to reveal statistical properties of the signum-generated
dichotomic sequence $\gamma_n$, one should associate its mean value
$\overline{\gamma}$, variance $C_\gamma(0)$ and pair correlator
$C_\gamma(r)$ with the corresponding independent characteristics,
namely, the mean value $\overline{\beta}$ and the modulation
function $G(n)$ [or, the same, with the intermediate correlator
$K_\beta(r)$]. According to the definition of average, one can
express the mean value $\overline{\gamma}$ in terms of
$\overline{\beta}$ via the error function~\cite{erf},
\begin{eqnarray}
\overline{\gamma}\equiv\overline{\gamma_n}&=& \int_{-\infty}^\infty
d\beta\,\rho_B(\beta)\,\mathrm{sign}(\beta)\nonumber\\[6pt]
&=&\sqrt{\frac{2}{\pi}}\int_{0}^{\overline{\beta}} dx\exp(-
x^2/2)\equiv\mathrm{erf}\big(\overline{\beta}/\sqrt{2}\big).
\label{MeanValue-gamma}
\end{eqnarray}

Similarly to Eq.~\eqref{Var-s}, the variance $C_{\gamma}(0)$ is
written in the form
\begin{equation}\label{Var-gamma}
C_\gamma(0)\equiv\overline{\gamma_n^2}-\overline{\gamma}^{\,2}
=1-\overline{\gamma}^{\,2}.
\end{equation}

An important characteristic of the stochastic sequence $\gamma_n$ is
the correlation function $C_{\gamma}(r)$,
\begin{equation}\label{C-gamma}
C_{\gamma}(r)\equiv\overline{\gamma_{n}\gamma_{n+r}}-\overline{\gamma}^2=
C_{\gamma}(0)K_{\gamma}(r).
\end{equation}
Its calculation is performed in Appendix~\ref{app-CorrBeta}. Here we
write down only the final equation that relates the correlator
$K_{\gamma}(r)$ to $K_\beta(r)$,
\begin{eqnarray}\label{Corr}
(1&-&\overline{\gamma}^2)K_{\gamma}(r)\nonumber\\[6pt]
&=&\frac{2}{\pi}\int_{0}^{K_\beta(r)} \frac{dx}{\sqrt{1-x^2}}
\exp\Big(-\frac{\overline{\beta}^2}{1+x}\Big).
\end{eqnarray}
Note that r.h.s. of the latter equation is not elementary function,
therefore its analytical study is not simple. However, the case
$\overline{\beta}=0$ allows one to perform complete analytical
analysis.

%%%%%%%%%%%%%%%%%%%%%%%%%%%%%%%%%%%%%%%%%%%%%%%%%%%%%%%%%%%%%%%%%%
\section{Unbiased Sequence}
\label{sec-UnbSeq}
%%%%%%%%%%%%%%%%%%%%%%%%%%%%%%%%%%%%%%%%%%%%%%%%%%%%%%%%%%%%%%%%%%

If the mean value of intermediate chain $\beta_n$ vanishes,
$\overline{\beta}=0$, then due to Eq.~\eqref{MeanValue-gamma}, the
mean value of generated sequence $\gamma_n$ vanishes also,
$\overline{\gamma}=0$. In this case the relation \eqref{Corr} turns
out to be remarkably simplified,
\begin{equation}\label{UnbSeq-Corr1}
K_{\gamma}(r)=\frac{2}{\pi}\int_{0}^{K_\beta(r)}
\frac{dx}{\sqrt{1-x^2}}=\frac{2}{\pi}\arcsin[K_\beta(r)].
\end{equation}
Another equivalent form is
\begin{equation}\label{UnbSeq-Corr2}
K_\beta(r)=\sin\Big[\frac{\pi}{2}K_{\gamma}(r)\Big].
\end{equation}
From the above relations one can conclude that the $\gamma$-sequence
generated with the discussed signum function method, is random.
Indeed, the decay of correlations with an increase of $|r|$ in the
intermediate $\beta$-chain, also results in the decay of
correlations in the generated dichotomic $\gamma$-sequence.

Now it is suitable to rewrite Eq.~\eqref{UnbSeq-Corr2} in the
Fourier representation,
\begin{equation}\label{FTr-Corr2}
\mathcal{K}_\beta(k)=\mathcal{S}\{K_{\gamma}\}(k).
\end{equation}
Here the symbol $\mathcal{S}\{\cdot\}(k)$ stands for the operator
that transforms the function $K(r)$ by the following rule,
\begin{subequations}\label{SofK}
\begin{eqnarray}
&&\mathcal{S}\{K\}(k)\equiv\sum_{r=-\infty}^\infty
\sin\Big[\frac{\pi}{2}K(r)\Big]\exp(-ikr)\label{SofK-def}\\[6pt]
&&=\Big(1-\frac{\pi}{2}\Big)+\frac{\pi}{2}\mathcal{K}(k)\nonumber\\[6pt]
&&+2\sum_{r=1}^\infty \left\{\sin\Big[\frac{\pi}{2}K(r)\Big]
-\frac{\pi}{2}K(r)\right\}\cos(kr).\label{SofK-num}
\end{eqnarray}
\end{subequations}
It is important to note that the series \eqref{SofK-def} can
converge very slowly. Therefore, in the analytical and numerical
analysis one has to take into account a lot of terms in the sum in
order to obtain correct result. To avoid this problem, we have used
the following trick that is based on the second equality
\eqref{SofK-num}. Namely, since $K(r)\to 0$ when $|r|\to\infty$, the
latter sum converges quite rapidly according to the asymptotic
relation
\begin{equation}%\label{}
\sin\Big[\frac{\pi}{2}K(r)\Big]-\frac{\pi}{2}K(r)
\to\frac{\pi^3}{48}\,K^3(r),\qquad|r|\to\infty\,.
\end{equation}

The substitution of Eq.~\eqref{FTr-Corr2} into Eq.~\eqref{G} yields
the following final relation between the modulation function $G(n)$
and the correlator $K_\gamma(r)$ of the generating dichotomic noise
$\gamma_n$,
\begin{equation}\label{UnbSeq-GK}
G(n)=\frac{1}{\pi}\int_{0}^{\pi}dk\,
\sqrt{\mathcal{S}\{K_{\gamma}\}(k)}\,\cos(kn).
\end{equation}
Since the correlator $K_\gamma(r)$ is supposed to be known, the
relation \eqref{UnbSeq-GK} should be regarded as the expression
determining the modulation function $G(n)$.

As one can see, the SFG method for constructing the correlated
dichotomic sequence $\gamma_n$ with the zero mean, unit variance and
prescribed two-point correlator $K_\gamma(r)$ reduces to the
following steps. First, starting from a desirable profile of
$K_\gamma(r)$ and employing Eqs.~\eqref{SofK} and \eqref{UnbSeq-GK},
one has to obtain the modulation function $G(n)$. After, the
correlated sequence $\gamma_n$ can be generated in accordance with
Eq.~\eqref{gamma-beta}. However, it is important to take into
account the restriction that directly follows from
Eq.~\eqref{FTr-Corr2}. Specifically, since the power spectrum of any
random process, in particular $\mathcal{K}_\beta(k)$, is a
non-negative function of the wave number $k$, the function
$\mathcal{S}\{K_{\gamma}\}(k)$ also has to be non-negative,
\begin{equation}\label{SKgamma>0}
\mathcal{S}\{K_{\gamma}\}(k)\geq0\qquad\mbox{for}\qquad
|k|\leqslant\pi.
\end{equation}
This condition becomes apparent from Eq.~\eqref{UnbSeq-GK}, in which
the function $\mathcal{S}\{K_{\gamma}\}(k)$ enters as a radicand. In
other words, with this method the function $K_{\gamma}(r)$ can be
considered as a correlator of a dichotomic random sequence
$\gamma_n$, if and only if $\mathcal{S}\{K_{\gamma}\}(k)$ is a
non-negative function of $k$.

In view of the revealed restriction \eqref{SKgamma>0}, let us make a
qualitative analysis of Eq.~\eqref{SofK-num}. Note that the first
summand $(1-\pi/2)$ is negative. Therefore, if the spectrum
$\mathcal{K}(k)$ vanishes within some interval of $k$ and the whole
function $\mathcal{S}\{K\}(k)$ is non-negative there, consequently,
the third summand is positive, exceeding the value $(\pi/2-1)$.
However, in the third summand only first few terms determine its
sign and give correct estimate of the magnitude. Thus, one may
expect that the correlation function of a dichotomic chain, which
spectrum is zero in some interval, cannot be generated by the SFG
method.

\begin{figure}[htbp!]
\scalebox{0.8}[0.8]{\includegraphics{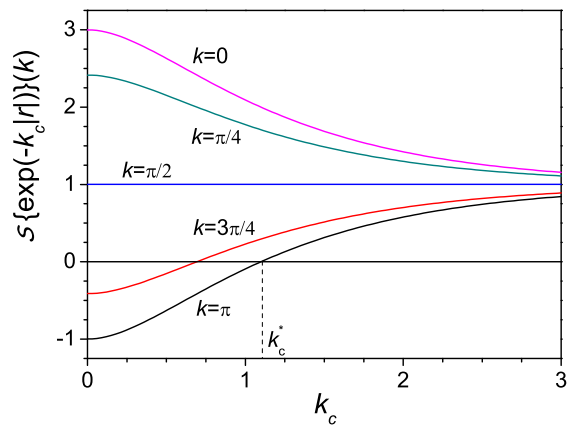}}
\caption{\label{S_of_kc} (Color online) Dependence
$\mathcal{S}\big\{\exp(-k_c|r|)\big\}(k)$ on the correlation
parameter $k_c$ for several values of wave number $k$.}
\end{figure}

It should be stressed that the condition \eqref{SKgamma>0} is a
quite serious restriction to the type of correlators allowed for
dichotomic sequences. As a demonstration, let us consider simple
example of the one-step additive Markov chain of variables
$\varepsilon_n=\{0,1\}$ obtained according to the conditional
probability
\begin{equation}\label{MarkovCondProb}
P(\varepsilon_n=1|\varepsilon_{n-1})=\overline{\varepsilon}+
(\varepsilon_{n-1}-\overline{\varepsilon})\exp(-k_c),
\end{equation}
where the parameter $k_c$ is the inverse correlation length. As is
known~\cite{IzKrMakMUYa06}, this chain has an exponential pair
correlator, $K_{\exp}(r)=\exp\big(-k_c|r|\big)$. It is instructive
that the function $\mathcal{S}\{K_{\exp}\}(k)$ takes negative values
for small $k_c$ below $k_c^*\approx1.099\ldots$, see the data in
Fig.~\ref{S_of_kc}. Thus, the SFG method can reproduce exponential
correlator only for $k_c\geqslant k_c^*$, however, not for small
$k_c$ in the most interesting region of long-range correlations.

%%%%%%%%%%%%%%%%%%%%%%%%%%%%%%%%%%%%%%%%%%%%%%%%%%%%%%%%%%%%%%%%%%
\section{Long-Range Correlators with Step-Wise Spectrum}
\label{sec-AnalyzeSW}
%%%%%%%%%%%%%%%%%%%%%%%%%%%%%%%%%%%%%%%%%%%%%%%%%%%%%%%%%%%%%%%%%%

%%%%%%%%%%%%%%%%%%%%%%%%%%%%%%%%%%%%%%%%%%%%%%%%%%%%%%%%%%%%%%%%%%
\subsection{Maximal Jump}
\label{subsec-SWinf}
%%%%%%%%%%%%%%%%%%%%%%%%%%%%%%%%%%%%%%%%%%%%%%%%%%%%%%%%%%%%%%%%%%

Here we demonstrate that the discussed method cannot be applied for
a construction of dichotomic sequences with long-range correlators
resulting in the step-wise power spectrum
\begin{subequations}\label{SW-K}
\begin{eqnarray}
K_\gamma(r)&=&\frac{\sin(k_cr)}{k_cr}\,,\label{SW-Kr}\\[6pt]
\mathcal{K}_\gamma(k)&=&\frac{\pi}{k_c}\Theta(k_c-|k|)\,,\quad
0<k_c\leqslant\pi,\quad|k|\leqslant\pi.\label{SW-Kk}
\end{eqnarray}
\end{subequations}
This kind of correlations is of specific interest in view of
applications to 1D disordered superlattices with a selective
transport, see, e.g. \cite{IzKr99,IzMak05}. Here $k_c$ is the
correlation parameter (inverse correlation length) to be specified,
and $\Theta(x)$ implies the Heaviside unit-step function,
$\Theta(x<0)=0$ and $\Theta(x>0)=1$.

\begin{figure}[htbp!]
\scalebox{0.75}[0.8]{\includegraphics{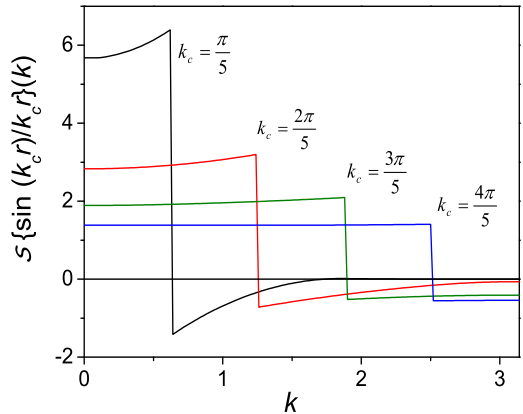}}
\caption{\label{examp_of_sin} (Color online) Dependence
$\mathcal{S}\big\{\sin (k_cr)/k_cr\big\}(k)$ on $k$ for different
values of $k_c$.}
\end{figure}

First, we analyze the values of $k_c$ within the interval
$0<k_c<\pi$. Two specific cases of $k_c=0$ and $k_c=\pi$ will be
considered afterwards.

In the analysis of the SFG method the crucial characteristic is the
function $\mathcal{S}\{K_{\gamma}\}(k)$ defined by Eq.~\eqref{SofK}.
It has to be non-negative for any value of the argument $k$ within
the interval $|k|\leqslant\pi$, see Eq.~\eqref{SKgamma>0}. In the
case of the long-range correlator \eqref{SW-Kr} it is suitable to
use the following explicit expression for
$\mathcal{S}\{K_{\gamma}\}(k)$,
\begin{eqnarray}\label{SW-S}
&&\mathcal{S}\left\{\frac{\sin(k_cr)}{k_cr}\right\}(k)=
\Big(1-\frac{\pi}{2}\Big)+\frac{\pi^2}{2k_c}\Theta(k_c-|k|)\nonumber\\[6pt]
&&+2\sum_{r=1}^\infty
\Big\{\sin\Big[\frac{\pi}{2}\frac{\sin(k_cr)}{k_cr}\Big]
-\frac{\pi}{2}\frac{\sin(k_cr)}{k_cr}\Big\}\cos(kr).
\end{eqnarray}
It is remarkable that the summand in the last term behaves as
$\pi^3/24k_c^3r^3$ when $r\to\infty$. Hence, at finite $k_c$ the sum
converges quite rapidly and \emph{uniformly}. Therefore, the sum is
a \emph{continuous} function of $k$, in particular, at $k=k_c$, and
can be easily calculated numerically.

\begin{figure}[htbp!]
\scalebox{0.75}[0.8]{\includegraphics{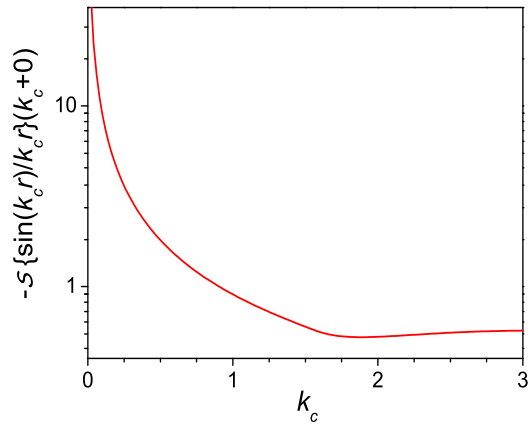}}
\caption{\label{S_of_kc+0} (Color online) Dependence
$-\mathcal{S}\big\{\sin(k_cr)/k_cr\big\}(k_c+0)$ vs $k_c$ in
log-scale. }
\end{figure}

In Fig.~\ref{examp_of_sin} the behavior of the radicand \eqref{SW-S}
in Eq.~\eqref{UnbSeq-GK} is shown for several values of the inverse
correlation length $k_c$. Since $\mathcal{S}\{\cdot\}(k)$ is an even
function of the wave number $k$, the discussion can be restricted by
the interval $0\leqslant k\leqslant\pi$. From Eq.~\eqref{SW-S} and
Fig.~\ref{examp_of_sin} one can draw the following conclusions.
\begin{enumerate}
\item
Due to the last term in expression \eqref{SW-S} the function
$\mathcal{S}\big\{\sin(k_cr)/k_cr\big\}(k)$ increases with an
increase of $k$ for all $k_c$ within both intervals $(0,k_c)$ and
$(k_c,\pi)$.

\item
The negative jump of $\mathcal{S}\big\{\sin(k_cr)/k_cr\big\}(k)$
occurs at $k=k_c$, at the same point where the power spectrum
\eqref{SW-Kk} has a jump. The maximal and minimal values of the
function are achieved at $k=k_c-0$ and $k=k_c+0$, respectively. The
jump is exclusively formed by the second term in Eq.~\eqref{SW-S}.
Therefore, its value reads
\begin{eqnarray}\label{SW-Jump}
&&\mathcal{S}\big\{\sin (k_cr)/k_cr\big\}(k=k_c-0)\nonumber\\[6pt]
&&-\mathcal{S}\big\{\sin (k_cr)/k_cr\big\}(k=k_c+0)=\pi^2/2k_c.
\end{eqnarray}

\item \label{negMin}
The minimal value $\mathcal{S}\big\{\sin (k_cr)/k_cr\big\}(k=k_c+0)$
is negative for all finite values of $k_c$ within the interval
$0<k_c<\pi$. The complete dependence of positive function
$-\mathcal{S}\big\{\sin (k_cr)/k_cr\big\}(k_c+0)$ on $k_c$ is
depicted in Fig.~\ref{S_of_kc+0}.
\end{enumerate}
Item~\ref{negMin} displays clearly that for all finite values
$k_c<\pi$ the correlator and corresponding power spectrum
\eqref{SW-K} cannot be created by making use of the discussed
method. Indeed, the modulation function $G(n)$ turns out to be of
complex value, see Eq.~\eqref{UnbSeq-GK}.

Now we determine the values of $k_c$ for which the function
\eqref{SW-Kr} can not be the pair correlator of a dichotomic
sequence regardless of the method of generation. This can be done
with the use of the results of Sec.~\ref{sec-NecCond}.

First, we demonstrate analytically that this function cannot be the
correlation function of a dichotomic sequence $s_{n}$ with arbitrary
mean value $\overline{s}$ for small but finite values of $k_c$,
namely, for $0<k_c\ll1$. To this end, we take Eq.~\eqref{norm_in1}
at $r'=1$ and $r=r_a=[a/k_c]$, where $[x]$ is the integer part of
$x$ and $a$ is a constant,
\begin{equation}%\label{}
\Big|\frac{\sin k_c}{k_c}-\frac{\sin k_c(r_a-1)}{k_c(r_a-1)}\Big|+
\frac{\sin k_cr_a}{k_cr_a}\leqslant1.
\end{equation}
Being expanded in small parameter $k_c$, this condition reads
\begin{equation}\label{SW-NCond1}
\Big(\frac{\cos a}{a}-\frac{\sin a}{a^2}\Big)k_c+O(k_c^2)\leqslant0.
\end{equation}
It is evident that the l.h.s. of Eq.~\eqref{SW-NCond1} can be
positive at some values of $a$ (e.g., for $a=2\pi+\pi/4$). Thus, the
requirement \eqref{norm_in1} is violated.

The inequality \eqref{norm_in2} can also result in new restrictions
for allowed values of $k_c$ for the case $\overline{s}=0$. Rewriting
it at $r=2$ and $r'=1$, we get
\begin{equation}\label{SW-NCond2}
4\sin k_c- \sin2k_c\leqslant2k_c.
\end{equation}
Numerical analysis shows that this condition does not hold true for
all finite $k_c$ from the interval $0<k_c< k^*$, where $k^*\approx
2.139\ldots$. It is met only at $k_c> k^*$. However, since
Eq.~\eqref{norm_in2} is just necessary condition, one cannot
guarantee an existence of the correlator \eqref{SW-Kr} even at $k_c>
k^*$.

At the critical point $k_c=\pi$ the correlator and spectrum
\eqref{SW-K} reduce to $K_\gamma(r)=\delta_{r,0}$ and
$\mathcal{K}_\gamma(k)=1$. This gives rise to the relations,
$\mathcal{S}\big\{\delta_{r,0}\big\}(k)=1$ and $G(n)=\delta_{n,0}$,
hence, $\beta_n=\alpha_n$. Consequently, one can apply the SFG.
However, this specific case of $k_c=\pi$ is not interesting because
from the Gaussian white-noise $\alpha_n$-sequence the method
fabricates the dichotomic chain $\gamma_n$ again of white-noise
type.

As to the singular point $k_c=0$, here we have $K_\gamma(r)=1$ for
the correlator, and $\mathcal{K}_\gamma(k)=2\pi\delta(k)$ for the
power spectrum, therefore, the radicand is
$\mathcal{S}\big\{1\big\}(k)=2\pi\delta(k)$. One can see that the
SFG is formally applicable. Besides, the necessary conditions
\eqref{SW-NCond1} and \eqref{SW-NCond2} are satisfied automatically.
However, this case is a singular one since for any arbitrarily small
but finite values of $k_c$ it is not possible to create a dichotomic
sequence with the correlation properties \eqref{SW-K} neither by the
SFG or by any other method.

%%%%%%%%%%%%%%%%%%%%%%%%%%%%%%%%%%%%%%%%%%%%%%%%%%%%%%%%%%%%%%%%%%
\subsection{Partial Jump}
\label{subsec-SWfin}
%%%%%%%%%%%%%%%%%%%%%%%%%%%%%%%%%%%%%%%%%%%%%%%%%%%%%%%%%%%%%%%%%%

Now we extend our analysis to a more general correlation function
that may have various applications. This function also results in a
step-wise power spectrum, however, with an additional parameter $h$
that determines the height of step,
\begin{subequations}\label{SW-Kh}
\begin{eqnarray}
K_{\gamma,h}(r)&=&h\,\delta_{r,0}+(1-h)\frac{\sin(k_cr)}{k_cr}\,,
\label{SW-Khr}\\[6pt]
\mathcal{K}_{\gamma,h}(k)&=&h+(1-h)\frac{\pi}{k_c}\Theta(k_c-|k|)>0\,,
\label{SW-Khk}\\[6pt]
&&0\leqslant h\leqslant1,
\qquad0<k_c\leqslant\pi,\quad|k|\leqslant\pi.\nonumber
\end{eqnarray}
\end{subequations}
Eq.~\eqref{SW-Kh} coincides with Eq.~\eqref{SW-K} if the
step-parameter $h=0$. Otherwise, when $h=1$ the generated
$\gamma$-sequence turns into a dichotomic white noise independently
of $k_c$. Also, $\gamma_n$ becomes delta-correlated at $k_c=\pi$ for
arbitrary $h$. Therefore, at $k_c=\pi$ the conclusions of the
previous subsection are also valid.

In what follows, it is convenient to analyze finite values of
$k_c<\pi$. The power spectrum \eqref{SW-Khk} is an even function of
the wave number $k$ and has two symmetric jumps at the points $k=\pm
k_c$. For positive $0<k\leqslant\pi$ the spectrum abruptly falls
down at $k=k_c$ from the maximal value
$\mathcal{K}_{\gamma,h}(k<k_c)=h+(1-h)\pi/k_c$ to the minimal one,
$\mathcal{K}_{\gamma,h}(k>k_c)=h$. Evidently, this jump can be
regarded as a mobility edge of disordered 1D conductors, if the
maximal value $h+(1-h)\pi/k_c$ is much larger than the minimal one,
$h$,
\begin{equation}\label{MobEdge}
1+\frac{1-h}{h}\,\frac{\pi}{k_c}\gg1.
\end{equation}
One can see that for finite $h$ well above zero, this is possible
only for small $k_c$. What is more tricky, for $k_c\ll1$ the
mobility edge may emerge even in the case when the generated
$\gamma$-sequence is close to a dichotomic white noise, i.e., when
$1-h\ll1$. Therefore, one should have,
\begin{equation}\label{MobEdgeWN}
0<k_c\ll1-h\ll1.
\end{equation}
The reason of existence of the mobility edge under the conditions
\eqref{MobEdgeWN} is a significant contribution of the second term
in the coorelator \eqref{SW-Khr}. In spite of the fact that it has a
quite small amplitude $1-h\ll1$, this term provides extremely
long-range correlations with the characteristic scale
$k_c^{-1}\gg(1-h)^{-1}\gg1$.

Now we address the function $\mathcal{S}\big\{K_{\gamma,h}\big\}(k)$
that must be non-negative in order to construct the correlated
sequence $\gamma_n$ with the use of the SFG method. In accordance
with the definition \eqref{SofK}, an appropriate analysis can be
done with the following explicit expression,
\begin{eqnarray}\label{SWh-S}
\mathcal{S}\{K_{\gamma,h}\}(k)&=&1-\frac{\pi}{2}(1-h)
+(1-h)\frac{\pi^2}{2k_c}\Theta(k_c-|k|)\nonumber\\[6pt]
&&+2\sum_{r=1}^\infty
\Big\{\sin\Big[\frac{\pi}{2}(1-h)\frac{\sin(k_cr)}{k_cr}\Big]
\nonumber\\[6pt]
&&-\frac{\pi}{2}(1-h)\frac{\sin(k_cr)}{k_cr}\Big\}\cos(kr).
\end{eqnarray}
As in the previous case \eqref{SW-K}, the summand in the last term
behaves as $\pi^3/24k_c^3r^3$ if $r\to\infty$. Hence, at finite
$k_c$ the sum converges rapidly and \emph{uniformly}. It is a
\emph{continuous} function of $k$ , in particular, at $k=k_c$.

\begin{figure}[htbp!]
\scalebox{0.75}[0.8]{\includegraphics{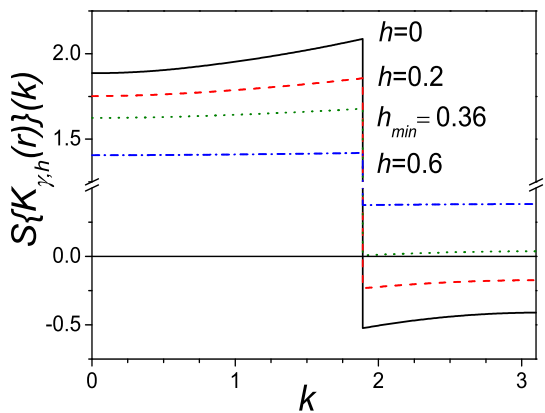}}
\caption{\label{W_HMIN} (Color online) Function
$\mathcal{S}\big\{K_{\gamma,h}(r)\}(k)$ vs $k$ for few values of $h$
at $k_c=1.89$ (this $k_c$ is very close to the minimum point of the
curve in Fig.~\ref{S_of_kc+0} and to the minimum point of the upper
curve in Fig.~\ref{h_of_kc}). The function is entirely positive if
$h>h_{min}$. When $h=h_{min}$, the function goes to zero solely at
one point $k=k_c+0$ and is positive otherwise. For $h<h_{min}$, the
function is negative within the whole interval $k_c<k<\pi$.}
\end{figure}

The numerical calculations of
$\mathcal{S}\big\{K_{\gamma,h}(r)\big\}(k)$ performed for finite
$0<k_c<\pi$ and $0\leqslant k\leqslant\pi$, are shown in
Fig.~\ref{W_HMIN}. Together with Eq.~\eqref{SWh-S} they provide us
with the following empirical results.
\begin{enumerate}
\item
Due to the last term in Eq.~\eqref{SWh-S}, the function
$\mathcal{S}\big\{K_{\gamma,h}\big\}(k)$ increases with $k$ at
arbitrary values of $k_c$ and $h$ within both intervals $(0,k_c)$
and $(k_c,\pi)$.

\item
The negative jump of $\mathcal{S}\big\{K_{\gamma,h}\big\}(k)$ occurs
at the same point $k=k_c$ as for the jump of the prescribed power
spectrum \eqref{SW-Khk}. The maximal and minimal values of the
function are reached, respectively, at $k=k_c-0$ and $k=k_c+0$ for
all values of the step-parameter $h$ within $0\leqslant h<1$. The
jump is exclusively related to the third term in Eq.~\eqref{SWh-S}.
Therefore, its value reads
\begin{equation}\label{SWh-Jump}
\mathcal{S}\big\{K_{\gamma,h}\big\}(k_c-0)-
\mathcal{S}\big\{K_{\gamma,h}\big\}(k_c+0)=(1-h)\frac{\pi^2}{2k_c}.
\end{equation}

\item
Depending on $h$, the minimum
$\mathcal{S}\big\{K_{\gamma,h}\big\}(k_c+0)$ can be either negative
or positive. Also, the value of
$\mathcal{S}\big\{K_{\gamma,h}\big\}(k_c+0)$ monotonically increases
with an increase of $h$.
\end{enumerate}

The presented numerical analysis can be supplemented with the
following two points. First, from the treatment of the case
\eqref{SW-K}, we know that
\begin{equation}\label{Sminh0}
\mathcal{S}\big\{K_{\gamma,h}\big\}(k_c+0)<0\qquad\mbox{for}\quad
h=0.
\end{equation}
Second, from the definition \eqref{SWh-S} one can easily reveal that
\begin{equation}\label{Sh1}
\mathcal{S}\big\{K_{\gamma,h}\big\}(k)=1\qquad\mbox{for}\quad h=1.
\end{equation}

\begin{figure}[htbp!]
\scalebox{0.75}[0.8]{\includegraphics{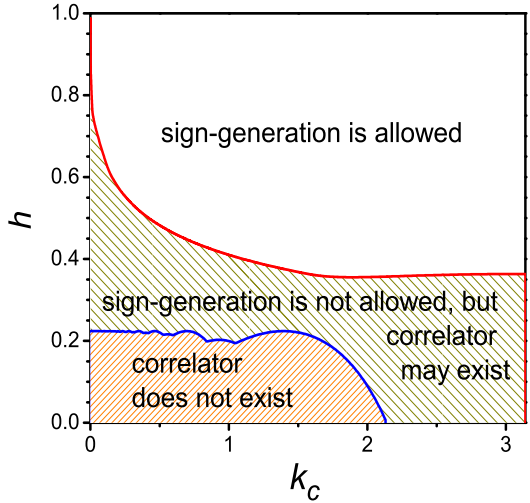}}
\caption{\label{h_of_kc} (Color online) Space of control parameters
$k_c$ and $h$. Upper curve is the dependence $h_{min}(k_c)$ while
low curve depicts $h_0(k_c)$. Within the lowest area in which
$0\leqslant h<h_0$, a dichotomic sequence $\gamma_n$ with the
correlator $K_{\gamma,h}(r)$ does not exist. The area with
$h_{min}\leqslant h\leqslant1$ allows for a creation of $\gamma_n$
with $K_{\gamma,h}(r)$ by the discussed method. In the intermediate
region $h_0\leqslant h<h_{min}$ the SFG method does not work, and an
existence of a dichotomic sequence with the step-wise spectrum
remains an open problem.}
\end{figure}

Summarizing our results, one can conclude that there exists a
threshold value of the step-parameter $h$ that we refer to as
$h_{min}$, that separates the region $0\leqslant h<h_{min}$ in which
$\mathcal{S}\big\{K_{\gamma,h}\big\}(k)$ has negative values, from
the region with non-negative values,
\begin{equation}\label{hmin}
\mathcal{S}\big\{K_{\gamma,h}\big\}(k)\geqslant0\qquad \mbox{for}
\quad h_{min}\leqslant h\leqslant1.
\end{equation}
It is clear that the threshold $h_{min}$ obeys the equation (see
Fig.~\ref{W_HMIN})
\begin{equation}\label{hmin-def}
\mathcal{S}\big\{K_{\gamma,h_{min}}\big\}(k_c+0)=0,
\end{equation}
and depends on the correlation parameter $k_c$. By substitution of
Eq.~\eqref{SWh-S} into Eq.~\eqref{hmin-def}, one can rewrite it in
explicit form,
\begin{eqnarray}\label{hmin-eq}
&&\frac{\pi}{4}(1-h_{min})-\frac{1}{2}=\sum_{r=1}^\infty
\Big\{\sin\Big[\frac{\pi}{2}(1-h_{min})\dfrac{\sin(k_cr)}{k_cr}\Big]
\nonumber\\[6pt]
&&-\frac{\pi}{2}(1-h_{min})\dfrac{\sin(k_cr)}{k_cr}\Big\}\cos(k_cr).
\label{hmin_num}
\end{eqnarray}
The numerical solution $h_{min}(k_c)$ of this equation is displayed
in Fig.~\ref{h_of_kc}  by the upper curve.

Eq.~\eqref{hmin-eq} can be solved analytically at small $k_c$,
resulting in
\begin{equation}\label{hminkcsmall}
h_{min}(k_c)=1-\left(\frac{96k_c}{\pi^4}\right)^{1/3}+
O\big(k_c^{2/3}\big)\quad\mbox{for}\quad k_c\ll1.
\end{equation}
This expression exhibits the limit $h_{min}\to1$ for $k_c\to0$.

Thus, in accordance with the study performed above, the dichotomic
sequence $\gamma_n$ with long-range correlator and step-wise power
spectrum \eqref{SW-Kh} can be constructed by the SFG method only if
its parameters $k_c$ and $h$ are placed onto or above the upper line
in Fig.~\ref{h_of_kc}. Only in this area of the $(k_c,h)$-plane the
condition \eqref{hmin} holds true. Unfortunately, practically within
this whole area the parameters $k_c$ and $h$ have values of the
order of one, and, therefore, one cannot satisfy a quite strong
requirement \eqref{MobEdge} in order to clearly observe a mobility
edge. The only exception is a narrow vicinity of the point
$(k_c=0,h=1)$. Remarkably, in this vicinity due to specific
dependence \eqref{hminkcsmall} of $h_{min}(k_c)$, the conditions
\eqref{MobEdgeWN} can be satisfied,
\begin{equation}\label{MobEdgeSmallkc}
0<k_c\ll1-h\leqslant1-h_{min}(k_c)\approx
\left(\frac{96k_c}{\pi^4}\right)^{1/3}\ll1,
\end{equation}
and, consequently, the mobility edge can be achieved. So,
Eq.~\eqref{MobEdgeSmallkc} gives us the only (perhaps, just purely
theoretical) possibility to arrange a mobility edge in the transport
through the $\gamma$-sequence constructed by the SFG method.

Now we analyze the consequences of the necessary conditions
formulated in Sec.~\ref{sec-NecCond}. For the long-range correlator
\eqref{SW-Khr} the inequalities \eqref{normNecCond} lead to the
following restriction with respect to the step-parameter $h$,
\begin{equation}\label{SWh-NecCond}
1-\max{}^{-1}\left\{R(r,r')\right\}\leqslant h
\end{equation}
at $r\neq0$, $r'\neq0$ and $r\neq r'$. Here we have introduced the
function
\begin{equation}\label{R-def}
R(r,r')=\Big|\dfrac{\sin k_cr'}{k_cr'}\pm\dfrac{\sin
k_c(r-r')}{k_c(r-r')}\Big| \mp\dfrac{\sin k_cr}{k_cr}.
\end{equation}
Since by the definition $h\geqslant0$, the requirement
\eqref{SWh-NecCond} is meaningful only if its l.h.s. is positive.
Otherwise, it is satisfied automatically. The combination of
Eq.~\eqref{SWh-NecCond} with the assumption $0\leqslant h\leqslant1$
gives rise to the relation,
\begin{equation}\label{SWh-NecCondFin}
h_0\leqslant h\leqslant1,
\end{equation}
where new characteristic quantity $h_0$ is introduced,
\begin{equation}\label{SWh-h0}
h_0=1-\max{}^{-1}\left\{1,\max\{R(r,r')\}\right\}.
\end{equation}
This function $h_0(k_c)$ is shown in Fig.~\ref{h_of_kc} by low
curve. Its piecewise shape is caused by the fact that different $r$
and $r'$ contribute to $h_0$ within different intervals of $k_c$.
Finally, when $k_c$ becomes equal or larger than $k^*\backsimeq
2.139\ldots$ [see text after Eq.~\eqref{SW-NCond2}], we have $h_0=0$
and the necessary conditions \eqref{SWh-NecCondFin} reduce to the
initial ones, $0\leqslant h\leqslant1$.

Thus, taking into account that the required area of the parameters
$k_c$ and $h$ is determined by the relation, $0<k_c<\pi$,
$0\leqslant h\leqslant1$ of the $(k_c,h)$-plane, one can summarize
the following. A dichotomic sequence with long-range correlator and
step-wise power spectrum \eqref{SW-Kh} does not exist within the
lowest region $0\leqslant h<h_0(k_c)$. When $h_0(k_c)\leqslant
h<h_{min}(k_c)$, the dichotomic chain cannot be created by the SFG
method and there is no answer whether it can be created by any other
method. Finally, within the highest zone with $h_{min}(k_c)\leqslant
h\leqslant1$, one can construct desired dichotomic sequences with
the use of the discussed method.

%%%%%%%%%%%%%%%%%%%%%%%%%%%%%%%%%%%%%%%%%%%%%%%%%%%%%%%%%%%%%%%%%%
\subsection{Predefined Intermediate Spectrum}
\label{subsec-SWKbeta}
%%%%%%%%%%%%%%%%%%%%%%%%%%%%%%%%%%%%%%%%%%%%%%%%%%%%%%%%%%%%%%%%%%

With the SFG method, we first generate intermediate Gaussian
sequence $\beta_n$ and after, the dichotomic sequence $\gamma_n$.
Above, we specified the correlator $K_{\gamma}(r)$ of a final
dichotomic $\gamma$-sequence and analyzed the spectrum
$\mathcal{K}_\beta(k)=\mathcal{S}\{K_{\gamma}\}(k)$ of the
intermediate Gaussian $\beta$-chain, keeping in mind that the latter
must be non-negative, see Eqs.~\eqref{FTr-Corr2}, \eqref{UnbSeq-GK}.

Below, we ask question about the type of the spectrum of $\gamma_n$,
that emerges if the intermediate Gaussian sequence $\beta_n$ is
assumed to have given pair correlator with step-wise power spectrum
of the following form,
\begin{subequations}\label{SW-Kbeta}
\begin{eqnarray}
K_\beta(r)&=&\frac{\sin(k_cr)}{k_cr}\,,\label{SW-Kbetar}\\[6pt]
\mathcal{K}_\beta(k)&=&\frac{\pi}{k_c}\Theta(k_c-|k|)\,,\quad
0<k_c\leqslant\pi,\,|k|\leqslant\pi.\label{SW-Kbetak}
\end{eqnarray}
\end{subequations}
In accordance with the relation \eqref{UnbSeq-Corr1} and Fourier
transforms \eqref{FTr-K}, the corresponding correlator and power
spectrum of the dichotomic $\gamma$-sequence read,
\begin{subequations}\label{SW-Kgamma}
\begin{eqnarray}
K_\gamma(r)=\frac{2}{\pi}\arcsin\left[\frac{\sin(k_cr)}{k_cr}\right]\,,
\label{SW-Kgammar}\\[6pt]
\mathcal{K}_\gamma(k)=\frac{2}{\pi}\sum_{r=-\infty}^{\infty}
\arcsin\left[\frac{\sin(k_cr)}{k_cr}\right]\exp(-ikr)\nonumber\\[6pt]
=\Big(1-\frac{2}{\pi}\Big)+\frac{2}{k_c}\Theta(k_c-|k|)\nonumber\\[6pt]
+\frac{4}{\pi}\sum_{r=1}^\infty
\Big\{\arcsin\Big[\frac{\sin(k_cr)}{k_cr}\Big]
-\frac{\sin(k_cr)}{k_cr}\Big\}\cos(kr). \label{SW-Kgammak}
\end{eqnarray}
\end{subequations}

\begin{figure}[htbp!]
\scalebox{0.75}[0.8]{\includegraphics{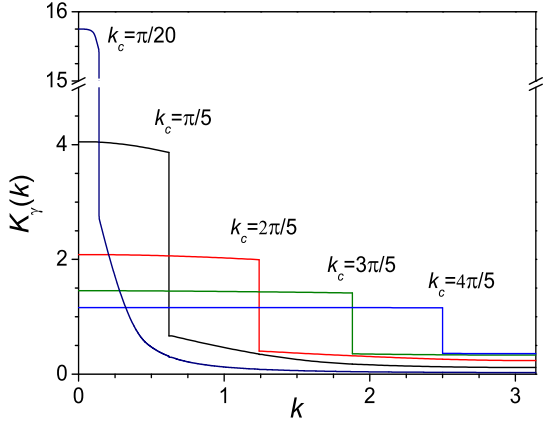}}
\caption{\label{Kgamma-arcsin} (Color online) Power spectrum
$\mathcal{K}_\gamma(k)$ vs $k$ for different values of the
correlation parameter $k_c$.}
\end{figure}

Here, last representation for the spectrum $\mathcal{K}_\gamma(k)$
is similar to that we have employed for the function
$\mathcal{S}\{\cdot\}(k)$ in its analysis [compare with
Eqs.~\eqref{SofK} and \eqref{SW-S}]. It is noteworthy to emphasize
that the summand in the last term of this representation behaves as
$2/3\pi k_c^3r^3$ when $r\to\infty$. Hence, as above, the sum
converges rapidly and \emph{uniformly}. Therefore, it can be easily
calculated numerically.

Fig.~\ref{Kgamma-arcsin} presents the behavior of the power spectrum
\eqref{SW-Kgammak} for dichotomic $\gamma_n$ in the case of the
predefined step-wise profile \eqref{SW-Kbetak} for the spectrum of
intermediate sequence $\beta_n$. Since $\mathcal{K}_\gamma(k)$ is an
even function of the wave number $k$, the presentation is sufficient
within $0\leqslant k\leqslant\pi$. Assuming $k_c<\pi$, from
Eq.~\eqref{SW-Kgammak} and Fig.~\ref{Kgamma-arcsin} one can
conclude:
\begin{enumerate}
\item
The function \eqref{SW-Kgammak} is positive within the whole
interval $|k|\leqslant\pi$. Therefore, it truly serves as a power
spectrum and its inverse Fourier transform \eqref{SW-Kgammar} is
valid correlator of the generated sequence $\gamma_n$;

\item
Due to the second and last terms in Eq.~\eqref{SW-Kgammak} the
spectrum $\mathcal{K}_\gamma(k)$ decreases with an increase of $k$
for all $k_c$ within whole interval $(0,\pi)$;

\item
The negative jump of $\mathcal{K}_\gamma(k)$ occurs at $k=k_c$, at
the same point as for the jump of prescribed intermediate power
spectrum $\mathcal{K}_\beta(k)$. The maximal and minimal values of
$\mathcal{K}_\gamma(k)$ at the jump are defined at $k=k_c-0$ and
$k=k_c+0$, respectively. The jump is exclusively formed by the
second term in Eq.~\eqref{SW-Kgammak}. Therefore, its value reads
\begin{equation}\label{SW-Kgammak-Jump}
\mathcal{K}_\gamma(k_c-0)-\mathcal{K}_\gamma(k_c+0)=2/k_c;
\end{equation}

\item
The value $\mathcal{K}_\gamma(k=k_c-0)$ to the left from the jump is
always positive. The value $\mathcal{K}_\gamma(k=k_c+0)$ to its
right is also positive for all $k_c$. Their ratio
%%%%%%%%%%%%%%%%%%%%%%%%%%%%%%%%%%%   new part  %%%%%%%%%%%%%%%%%%%%%%%%   done
$\mathcal{K}_\gamma(k_c-0)/\mathcal{K}_\gamma(k_c+0)$ being of the
order of one for finite $k_c$, seems to slightly increase with a
decrease of $k_c$. However, it is saturated if the inverse
correlation length $k_c$ vanishes. Indeed, from
Eq.~\eqref{SW-Kgammak} one can easily get,
\begin{subequations}\label{SW-Kgammak-Ratio}
\begin{eqnarray}
\frac{\mathcal{K}_\gamma(k_c-0)}{\mathcal{K}_\gamma(k_c+0)}
&=&\frac{\pi+I(k_c)+(\pi/2-1)k_c}{I(k_c)+(\pi/2-1)k_c}
\label{SW-Kgammak-RatioGen}\\[6pt]
&\to&\frac{\pi+I(0)}{I(0)}=6.0096\dots\,\,\mbox{if}\,\,k_c\to0.
\label{SW-Kgammak-Ratiokc0}
\end{eqnarray}
\end{subequations}
Here we have introduced
\begin{subequations}\label{SW-Kgammak-I}
\begin{eqnarray}
I(k_c)&=&2k_c\sum_{r=1}^\infty
\Big\{\arcsin\Big[\frac{\sin(k_cr)}{k_cr}\Big] \nonumber\\[6pt]
&&-\frac{\sin(k_cr)}{k_cr}\Big\}\cos(k_cr);
\label{SW-Kgammak-Idef}\\[6pt]
I(0)&=& 2\int\limits_0^\infty dx \left\{\arcsin\Big[\frac{\sin
x}{x}\Big] -\frac{\sin x}{x}\right\}\cos x. \label{SW-Kgammak-Ikc0}
\end{eqnarray}
\end{subequations}
In spite of the divergence of the jump \eqref{SW-Kgammak-Jump}, the
convergence of the ratio \eqref{SW-Kgammak-Ratio} at $k_c\to0$
occurs because the values of the spectrum $\mathcal{K}_\gamma(k)$ to
the left, $\mathcal{K}_\gamma(k_c-0)$, and to the right,
$\mathcal{K}_\gamma(k_c+0)$, from the jump, increase with decreasing
of $k_c$ exactly in the same manner,
$\mathcal{K}_\gamma(k_c-0)\approx2[\pi+I(0)]/\pi k_c$ and
$\mathcal{K}_\gamma(k_c+0)\approx2I(0)/\pi k_c$.
\end{enumerate}

In addition, it can be analytically shown that in the limit
$k_c\to0$, the correlator $K_\gamma(r)$ tends to unity, while the
spectrum $\mathcal{K}_\gamma(k)$ turns into the Dirac
delta-function,
\begin{subequations}\label{SW-Kgamma-kc0}
\begin{eqnarray}
&&\lim_{k_c\to0}K_\gamma(r)=
\frac{2}{\pi}\lim_{k_c\to0}\arcsin\left[\frac{\sin(k_cr)}{k_cr}\right]=1\,;
\label{SW-Kgammar-kc0}\\[6pt]
&&\lim_{k_c\to0}\mathcal{K}_\gamma(k)=2\pi\delta(k).\label{SW-Kgammak-kc0}
\end{eqnarray}
\end{subequations}
Therefore, as the correlation parameter $k_c$ vanishes, the final
dichotomic $\gamma$-sequence becomes to have extremely long-range
correlations.

On the contrary, for $k_c=\pi$ the correlator $K_\gamma(r)$ reduces
to the Kronecker delta-symbol, whereas the power spectrum
$\mathcal{K}_\gamma(k)$ degenerates into unity,
\begin{subequations}\label{SW-Kgamma-kcpi}
\begin{eqnarray}
&&K_\gamma(r)=\frac{2}{\pi}\arcsin\left[\delta_{r,0}\right]
=\delta_{r,0}\,;\label{SW-Kgammar-kcpi}\\[6pt]
&&\mathcal{K}_\gamma(k)=1\qquad\mbox{for}\quad
k_c=\pi.\label{SW-Kgammak-kcpi}
\end{eqnarray}
\end{subequations}
Thus, the final dichotomic $\gamma_n$ reduces to the white-noise
chain.

In summary, in spite of the fact that the jump ratio
$\mathcal{K}_\gamma(k_c-0)/\mathcal{K}_\gamma(k_c+0)$ in the power
spectrum $\mathcal{K}_\gamma(k)$ is of the order of unity for all
$k_c<\pi$, we hope that the mobility edge can be observed at small
enough correlation parameter $k_c\ll1$, due to specific
delta-function behavior of $\mathcal{K}_\gamma(k)$ itself [see
Fig.~\ref{Kgamma-arcsin} and Eq.~\eqref{SW-Kgammak-kc0}].

%%%%%%%%%%%%%%%%%%%%%%%%%%%%%%%%%%%%%%%%%%%%%%%%%%%%%%%%%%%%%%%%%%
\section{Power Correlators}
\label{sec-PowerCorr}
%%%%%%%%%%%%%%%%%%%%%%%%%%%%%%%%%%%%%%%%%%%%%%%%%%%%%%%%%%%%%%%%%%

%%%%%%%%%%%%%%%%%%%%%%%%%%%%%%%%%%%%%%%%%%%%%%%%%%%%%%%%%%%%%%%%%%
\subsection{Power Correlator for Dichotomic Sequence}
\label{subsec-PCKgamma}
%%%%%%%%%%%%%%%%%%%%%%%%%%%%%%%%%%%%%%%%%%%%%%%%%%%%%%%%%%%%%%%%%%

Here we consider an important problem of constructing a sequence
with the power correlation function and corresponding spectrum,
\begin{subequations}\label{PC-K}
\begin{eqnarray}
K_{\gamma,p}(r)&=&\delta_{r,0}+(k_c|r|)^{-p}(1-\delta_{r,0}),
\label{PC-Kgammar}\\[6pt]
\mathcal{K}_{\gamma,p}(k)&=&1\nonumber\\[6pt]
&+&k_c^{-p}\Big\{\mathrm{Li}_{p}\big[\exp(ik)\big]
+\mathrm{Li}_{p}\big[\exp(-ik)\big]\Big\},\label{PC-Kgammak}\\[6pt]
&&p>0,\quad k_c\geqslant1,\quad |k|\leqslant\pi.\nonumber
\end{eqnarray}
\end{subequations}
Here $p$ and $k_c$ are positive real numbers characterizing how fast
the correlator decreases. Note that the parameter $k_c$ cannot be
less than one since $K_{\gamma,p}(r)\leqslant1$. The Fourier
transform $\mathcal{K}_{\gamma,p}(k)$ of this correlator is
expressed via the polylogarithm function $\mathrm{Li}_{q}(z)$ that
is defined by
\begin{equation}\label{polylog}
\mathrm{Li}_{q}(z)=\sum_{r=1}^\infty\frac{z^r}{r^q}.
\end{equation}

\begin{figure}[htbp!]
\scalebox{0.75}[0.8]{\includegraphics{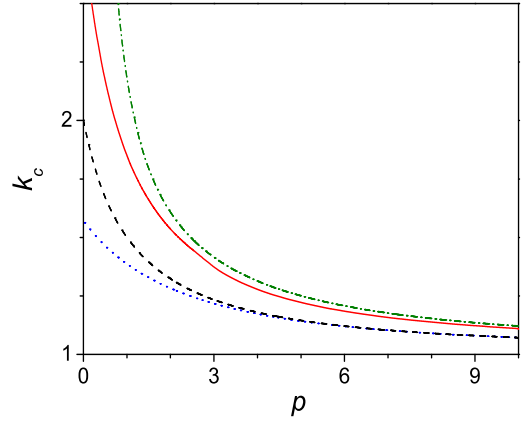}}
\caption{\label{Power} (Color online) {Various borders of the
parameters $p$ and $k_c$ above which the following relations are
fulfilled: (a) Eq.~\eqref{PC-KkPositive} (dotted curve), (b)
Eq.~\eqref{PC-SkPositive}) (solid curve), (c)
Eq.~\eqref{NecCondPow}) (dashed curve), (d) Eq.~\eqref{ApprPow1}
(dash-dotted curve).}}
\end{figure}

For $K_{\gamma,p}(r)$ to be the correlator of a stochastic process,
it is necessary to have $\mathcal{K}_{\gamma,p}(k)\geqslant0$ for
all $k$. This condition is satisfied if and only if the following
inequality is fulfilled,
\begin{equation}\label{PC-KkPositive}
k_c\geqslant\big[-2\mathrm{Li}_{p}(-1)\big]^{1/p}.
\end{equation}
This result is due to the shape of spectrum
$\mathcal{K}_{\gamma,p}(k)$ that monotonously decreases with an
increase of $k$ within the interval $(0,\pi)$, and reaches its
minimal value at $k=\pi$. The r.h.s. of the condition
\eqref{PC-KkPositive} can be calculated in limit cases,
\begin{eqnarray}\label{Li-limit}
\big[-2\mathrm{Li}_{p}(-1)\big]^{1/p}=\begin{cases}\pi/2-cp\,,& p\ll1,\\
1+p^{-1}\ln 2,& p\gg1\,,\end{cases}\\
c =  1.029\dots.\nonumber
\end{eqnarray}
In Fig.~\ref{Power} the area where power spectrum \eqref{PC-Kgammak}
is non-negative, is located above the dotted lowest curve.

If $K_{\gamma,p}(r)$ is the correlator of a \emph{dichotomic} random
sequence generated by the SFG method, then
$\mathcal{S}\{{K}_{\gamma,p}\}(k)\geqslant0$ for all values of $k$.
Since this function, as well as $\mathcal{K}_{\gamma,p}(k)$,
monotonously decreases, the only condition is required,
\begin{equation}\label{PC-SkPositive}
\mathcal{S}\{{K}_{\gamma,p}\}(\pi)\geqslant0.
\end{equation}
Expanding in Eq.~\eqref{SofK-def} the sin-function into series, we
get useful expression
\begin{eqnarray}\label{SCorrViaLi}
&&\mathcal{S}\{{K}_{\gamma,p}\}(k)=1+\sum_{l=0}^\infty
\frac{\pi^{2l+1}}{2^{2l+1}(2l+1)!}k_c^{-p(2l+1)}\nonumber\\
&&\times\Big\{\mathrm{Li}_{p(2l+1)}\big[\exp(ik)\big]
+\mathrm{Li}_{p(2l+1)}\big[\exp(-ik)\big]\Big\}.
\end{eqnarray}
For $p\to\infty$ it can be approximately calculated as
\begin{eqnarray}\label{SCorr-asymp}
\mathcal{S}\{{K}_{\gamma,p}\}(k)\backsimeq 1+2\sin\big(\pi
k_c^{-p}/2\big)\cos k.
\end{eqnarray}
Therefore, taking into account that $k_c\geqslant1$, one obtains the
following asymptotic of the condition \eqref{PC-SkPositive}
\begin{equation}\label{PC-SkPositive2}
k_c\geqslant1+p^{-1}\ln 3\qquad\mbox{for}\quad p\gg1.
\end{equation}
The area where the discussed SFG method is applicable located in
Fig.~\ref{Power} above solid curve.

Now we analyze the necessary condition for existence of the power
correlator \eqref{PC-Kgammar} for dichotomic sequence regardless of
the generation method. For the sake of simplicity we consider the
case $\overline{\gamma}=0$. From the
inequalities~\eqref{normNecCond} one can obtain the following
relation,
\begin{eqnarray}\label{PC-NecCond}
&&\max\left\{\big||r'|^{-p}\pm |r-r'|^{-p}\big|\mp|r|^{-p}\right\}
\leqslant k_c^p,
\\[6pt]
&&r\neq0,\quad r'\neq0,\quad r\neq r'.\nonumber
\end{eqnarray}
It is easy to see that the maximum with respect to $r'$, occurs at
$r'=\pm 1$ or $r'=r\pm 1$. Therefore, Eq.~\eqref{PC-NecCond} can be
rewritten as
\begin{equation}\label{PC-NecCond1}
\max_{r>0}\left\{1\pm (r+1)^{-p}\mp r^{-p}\right\}\leqslant k_c^p.
\end{equation}
The last condition is equivalent to
\begin{equation}\label{NecCondPow}
k_c\geqslant \big(2-2^{-p}\big)^{1/p}=\begin{cases}
2-2p\ln^22,&p\ll1,\\1+p^{-1}\ln2,&p\gg1.
\end{cases}
\end{equation}
In Fig.~\ref{Power} the border corresponding to this condition is
designated by dashed curve. It should be noted that
Eq.~\eqref{NecCondPow} is the necessary condition, thus, if it is
met, it is still not clear whether the correlator with such values
of parameters $p$ and $k_c$ exists.

%%%%%%%%%%%%%%%%%%%%%%%%%%%%%%%%%%%%%%%%%%%%%%%%%%%%%%%%%%%%%%%%%%
\subsection{Predefined Intermediate Power Correlator}
\label{subsec-PCKbeta}
%%%%%%%%%%%%%%%%%%%%%%%%%%%%%%%%%%%%%%%%%%%%%%%%%%%%%%%%%%%%%%%%%%

We have found that the dichotomic sequence $\gamma_n$ with the power
correlator \eqref{PC-Kgammar} can be constructed by the SFG method
for some values of parameters $p$ and $k_c$ that meet the condition
\eqref{PC-SkPositive}. Now let us take the intermediate Gaussian
sequence $\beta_n$ prescribed to have the power correlator and
corresponding spectrum
\begin{subequations}\label{PC-Kbeta}
\begin{eqnarray}
K_{\beta,p}(r)&=&\delta_{r,0}+(k_c|r|)^{-p}(1-\delta_{r,0}),
\label{PC-Kbetar}\\[6pt]
\mathcal{K}_{\beta,p}(k)&=&1\nonumber\\[6pt]
&+&k_c^{-p}\Big\{\mathrm{Li}_{p}\big[\exp(ik)\big]
+\mathrm{Li}_{p}\big[\exp(-ik)\big]\Big\},\label{PC-Kbetak}\\[6pt]
&&p>0,\quad k_c\geqslant1,\quad |k|\leqslant\pi.\nonumber
\end{eqnarray}
\end{subequations}
Evidently, the condition \eqref{PC-KkPositive} is implied to be met.
In accordance with the relation \eqref{UnbSeq-Corr1} and Fourier
transforms \eqref{FTr-K}, the correlator and power spectrum of the
generated dichotomic $\gamma$-sequence are described as
\begin{subequations}\label{IPC-Kgamma}
\begin{eqnarray}
K_{\gamma,p}(r)=\delta_{r,0}+\frac{2}{\pi}\arcsin
\big[(k_c|r|)^{-p}\big](1-\delta_{r,0}),
\label{IPC-Kgammar}\\[6pt]
\mathcal{K}_{\gamma,p}(k)=1+\frac{4}{\pi}\sum_{r=1}^\infty
\arcsin\Big[(k_cr)^{-p}\Big]\cos(kr). \label{IPC-Kgammak}
\end{eqnarray}
\end{subequations}
We can assert that since Eq.~\eqref{PC-KkPositive} is satisfied,
i.e. the spectrum \eqref{PC-Kbetak} of the intermediate
$\beta$-sequence is non-negative, then the spectrum
\eqref{IPC-Kgammak} of the generated dichotomic $\gamma_n$ is also
non-negative.

When $|r|\to \infty$, the correlator \eqref{IPC-Kgammar} tends to
zero in accordance with the following asymptotic
\begin{equation}\label{ApprPow}
K_{\gamma,p}(r)\backsimeq(k_c'|r|)^{-p},\qquad
k_c'=k_c(\pi/2)^{1/p}.
\end{equation}
As was shown, the allowed values of $k_c$ are expressed by
Eq.~\eqref{PC-KkPositive}. Therefore, the scaling parameter $k_c'$
should satisfy the condition
\begin{equation}\label{ApprPow1}
k_c'\geqslant\big[-\pi\mathrm{Li}_{p}(-1)\big]^{1/p}.
\end{equation}
This condition for possible values of $k_c'$ and $p$ is met in the
area above the dash-dotted curve in Fig.~\ref{Power}.

Thus, the mapping of the gaussian sequence with the power
correlation function \eqref{PC-Kbetar} into the binary sequence
result in the same power for the decrease of the final correlator
\eqref{IPC-Kgammar} expressed by Eq.~\eqref{ApprPow}. However, such
a behavior of the final correlator occurs only asymptotically, for
sufficiently large values of $|r|$.

%%%%%%%%%%%%%%%%%%%%%%%%%%%%%%%%%%%%%%%%%%%%%%%%%%%%%%%%%%%%%%%%%%
\section{Conclusion}
\label{app-Cnclsn}
%%%%%%%%%%%%%%%%%%%%%%%%%%%%%%%%%%%%%%%%%%%%%%%%%%%%%%%%%%%%%%%%%%

First, we would like to emphasize the following point that was
briefly mentioned in the beginning. Our study of the correlation
properties of a random dichotomic sequence $\gamma_n$ is based on
the example \eqref{BinSeqS} in which two elements are $``-1"$ and
$``1"$. On the other hand, there is a simple correspondence between
this chain and a dichotomic sequence $\varepsilon(n)$ consisting of
two arbitrary symbols $\varepsilon_0$ and $\varepsilon_1$,
\begin{equation}\label{BinSeqEpsilon}
\varepsilon(n)=\{\varepsilon_0,\varepsilon_1\},\qquad
n\in\textbf{\textbf{Z}} =\ldots,-2,-1,0,1,2,\ldots
\end{equation}
The correspondence is expressed by the linear relationship,
\begin{equation}\label{Epsilon-Gamma}
\varepsilon(n)=\frac{\varepsilon_0+\varepsilon_1}{2}\mp
\frac{\varepsilon_0-\varepsilon_1}{2}\,\gamma_n\,.
\end{equation}
The choice of the sign is not important. It derermines only into
what symbol, $\varepsilon_0$ or $\varepsilon_1$, the initial values
$``-1"$ and $``1"$ are converted.

In accordance with Eq.~\eqref{Epsilon-Gamma} and due to specific
properties \eqref{Var-gamma}, \eqref{s2=1} of the $\gamma$-sequence,
the connection between the mean values and variances is as follows,
\begin{subequations}\label{Eps-Gam-Avers}
\begin{eqnarray}
&&\varepsilon^2(n)=\frac{\varepsilon_0^2+\varepsilon_1^2}{2}\mp
\frac{\varepsilon_0^2-\varepsilon_1^2}{2}\,\gamma_n\,;
\label{Eps2-Gam}\\[6pt]
&&\overline{\varepsilon}=\frac{\varepsilon_0+\varepsilon_1}{2}\mp
\frac{\varepsilon_0-\varepsilon_1}{2}\,\overline{\gamma}\,;
\label{Eps-Gam-Aver}\\[6pt]
&&C_\varepsilon(0)\equiv\overline{\varepsilon^2(n)}-
\overline{\varepsilon}^2=\frac{(\varepsilon_0-\varepsilon_1)^2}{4}
C_\gamma(0)\,.\label{Eps-Gam-Var}
\end{eqnarray}
\end{subequations}
Analogously, the two-point correlation function $C_\varepsilon(r)$
of the $\varepsilon$-chain is associated with the binary correlation
function $C_\gamma(r)$ of the sequence $\gamma_n$ as follows
\begin{equation}\label{Eps-Gam-Corr}
C_\varepsilon(r)\equiv\overline{\varepsilon(n)\varepsilon(n+r)}-
\overline{\varepsilon}^2=\frac{(\varepsilon_0-\varepsilon_1)^2}{4}
C_\gamma(r)\,.
\end{equation}
The comparison of Eqs.~\eqref{Eps-Gam-Var} and \eqref{Eps-Gam-Corr}
makes obvious the equality between the normalized correlators
$K_\varepsilon(r)$ and $K_\gamma(r)$,
\begin{equation}\label{K=K}
K_\varepsilon(r)\equiv C_\varepsilon(r)/C_\varepsilon(0)=
C_\gamma(r)/C_\gamma(0)\equiv K_\gamma(r).
\end{equation}
Thus, our analysis is valid for any dichotomic sequence.

Our results can be summarized as follows. We have shown that the
statistical properties of random dichotomic sequences are
principally different from those known for sequences with a
continuous distribution of their elements. We were able to find
analytically the conditions \eqref{normNecCond} that can be used to
know whether a binary sequence can have the desired pair correlator.
Note that these two conditions are necessary only.

Another important restriction is due to the inequality
\eqref{SKgamma>0} derived under quite general assumptions. We have
shown that even in the well known case of an exponential decay of
correlations, there are no binary sequences that can be created with
the SFG method, unless the decay is sufficiently strong. This fact
is very important in view of many applications.

Our specific interest was in a possibility to create, with the
considered method, the binary sequences with long-range correlations
described by Eqs.~\eqref{SW-Kr} and \eqref{SW-Kk}. We have
analytically found that the function \eqref{SW-Kr} can not be a pair
correlator of any binary sequence. We have also examined a more
general correlation function [see Eq.~\eqref{SW-Kh}] that
corresponds to the generalization of the step-wise power spectrum.
Our extensive examination of the signum-function method, applied to
this correlation function, has revealed the regions of parameters
$k_c$ and $h$ for which the pair correlator can emerge in a binary
sequence. Correspondingly, we identified the regions where such a
pair correlator can not appear in binary sequences. For other values
of the control parameters we can not give definite answer,
therefore, a further study is needed.

Finally, we analyzed an important case of the power decay of the
pair correlator. Recently, the problem of the generation of random
processes with power correlations has attracted much attention in
the literature. Analyzing such correlators, we have found that the
SFG method in principle allows to construct binary sequences with
these correlators, however, with some restrictions on the values of
parameters in Eq.~\eqref{PC-K}.

\section{Acknowledgments} This work was partly supported by the
CONACYT (M\'exico) grant No~43730.

%%%%%%%%%%%
\appendix
%%%%%%%%%%%

%%%%%%%%%%%%%%%%%%%%%%%%%%%%%%%%%%%%%%%%%%%%%%%%%%%%%%%%%%%%%%%%%%
\section{Probability Density of $\beta_n$}
\label{app-RhoBeta}
%%%%%%%%%%%%%%%%%%%%%%%%%%%%%%%%%%%%%%%%%%%%%%%%%%%%%%%%%%%%%%%%%%

The standard way to derive the probability density $\rho_{B}(\beta)$
of the random process $\beta_n$ is due to its \emph{characteristic
function} $\varphi_{B}(t)$ defined by
\begin{equation}\label{ChF-def}
\varphi_{B}(t)\equiv\overline{\exp[it\beta_n]}=
\int_{-\infty}^{\infty}d\beta\,\rho_{B}(\beta)\exp(it\beta).
\end{equation}
From the last equality in this definition it immediately follows
that the probability density $\rho_{B}(\beta)$ is the Fourier
transform of $\varphi _{B}(t)$,
\begin{equation}\label{rhoB-phiB}
\rho_{B}(\beta)=\frac{1}{2\pi}\int_{-\infty }^{\infty }dt\,
\varphi_{B}(t)\exp(-it\beta).
\end{equation}

To start with, we substitute the explicit expression (\ref{beta})
for $\beta_n$ into the definition (\ref{ChF-def}) for characteristic
function $\varphi_{B}(t)$. Then, we rewrite the result as an
infinite product of exponential functions and take into account the
statistical independence of uncorrelated random variables
$\alpha_n$. This procedure yields
\begin{equation}\label{ChF-alpha}
\overline{\exp[it\beta_n]}= \exp(it\overline{\beta})
\prod_{n'=-\infty}^\infty\overline{\exp\big[itG(n-n')\alpha_{n'}\big]}.
\end{equation}

In accordance with the Gaussian distribution \eqref{rho-alpha} of
$\alpha_n$, its characteristic function is
\begin{eqnarray}
\varphi_A(\tau)\equiv\overline{\exp(i\tau\alpha_n)}&\equiv&
\int_{-\infty}^\infty d\alpha\,
\rho_{A}(\alpha)\exp(i\tau\alpha)\nonumber\\
&=&\exp(-\tau^2/2).\label{ChFofAlpha}
\end{eqnarray}

The use of Eqs.~\eqref{ChF-alpha}, \eqref{ChFofAlpha} with
$\tau=tG(n-n')$, and the normalization condition \eqref{G-Norm}
results in
\begin{equation}\label{ChFofBeta}
\varphi_B(t)=\exp(i\overline{\beta}t-t^2/2).
\end{equation}
As is known, this characteristic function corresponds to the
Gaussian probability density \eqref{rho-beta}. One can confirm this
fact by a direct evaluation of the integral in
Eq.~\eqref{rhoB-phiB}.

%%%%%%%%%%%%%%%%%%%%%%%%%%%%%%%%%%%%%%%%%%%%%%%%%%%%%%%%%%%%%%%%%%
\section{Pair Correlator of $\gamma_n$}
\label{app-CorrBeta}
%%%%%%%%%%%%%%%%%%%%%%%%%%%%%%%%%%%%%%%%%%%%%%%%%%%%%%%%%%%%%%%%%%

Let us derive the pair correlator $\overline{\gamma_n\gamma_{n+r}}$.
Employing the standard integral presentation for the signum
function,
\begin{equation}\label{sign}
\mathrm{sign}(z)=\frac{1}{\pi }\int_{-\infty}^\infty dx\,
\frac{\sin(zx)}{x}\,,
\end{equation}
and Eq.~\eqref{ChFofAlpha}, we arrive, in a manner similar to the
calculation of the characteristic function $\varphi_B(t)$ in
Appendix~\ref{app-RhoBeta}, at the expression,
\begin{eqnarray}\label{J-def}
&&\overline{\gamma_n\gamma_{n+r}}=
\mathcal{J}({K}_{\beta}(r),\overline{\beta})
\nonumber\\[6pt]
&&=\frac{2}{\pi^2} \int_{0}^\infty \frac{dx_1}{x_1}\int_{0}^\infty
\frac{dx_2}{x_2}\exp\Big(-\frac{x_1^2+x_2^2}2\Big)
\nonumber\\[6pt]
&&\times\sum_{s=-1,1}s\exp\big[sK_\beta(r)x_1x_2\big]
\cos\big[\overline{\beta}(x_1-sx_2)\big].
\end{eqnarray}

Thus, we have reduced the problem to the derivation of
$\mathcal{J}(K,\beta)$. To solve it, we obtain the derivative of
$\mathcal{J}(K,\beta)$ with respect to $K$. After some
simplifications one gets,
\begin{eqnarray}%\label{}
\frac{\partial}{\partial K}\mathcal{J}(K,\beta)=\frac{1}{\pi^2}
\int_{0}^\infty dx_1\int_{-\infty}^\infty
dx_2\exp\Big(-\frac{x_1^2+x_2^2}2\Big)
\nonumber\\[6pt]
\times\exp\big[Kx_1x_2\big]
\sum_{t=-1,1}\exp\big[it\beta(x_1-x_2)\big].\nonumber
\end{eqnarray}
To proceed, we write down the following relation that is valid for
arbitrary real quantities $a$ and $b$,
\begin{eqnarray}
\int_{-\infty}^\infty dx\exp(-x^2/2)\exp(ax)\exp(ibx)
\nonumber\\
=\sqrt{2\pi} \exp\big[(a^2-b^2)/2\big]\exp(iab).\label{int}
\end{eqnarray}

Using Eq.~\eqref{int} with $a=Kx_1$ and $b=-t\beta$ we integrate
over $x_2$ and  make further simplifications,
\begin{eqnarray}%\label{}
\notag \frac{\partial}{\partial K}\mathcal{J}(K,\beta)=
\frac{\sqrt{2\pi}}{\pi^2}  \int_{-\infty}^\infty
dx_1\exp\big[-x_1^2(1-K^2)/2\big]\\\notag \times
 \exp(-\beta^2/2) \exp\big[i\beta(1-K)x_1\big].
\end{eqnarray}
Now we change the integration variable $x_1$,
\begin{equation}%\label{}
x'_1=x_1\sqrt{1-K^2}.
\end{equation}
Then, applying Eq.~\eqref{int} with
\begin{equation}%\label{}
a=0,\qquad\qquad b=\beta\sqrt{\frac{1-K}{1+K}},
\end{equation}
we perform the integration over $x_1$ that gives rise to the
expression
\begin{eqnarray}\label{dJdK}
\frac{\partial}{\partial K}\mathcal{J}(K,\beta)=
\frac{2}{\pi\sqrt{1-K^2}} \exp\Big(-\frac{\beta^2}{1+K}\Big).
\end{eqnarray}

The general solution of Eq.~\eqref{dJdK} is
\begin{eqnarray}\label{J_J0+int}
\mathcal{J}(K,\beta)&=&\mathcal{J}(0,\beta)\nonumber\\
&+&\frac{2}{\pi}\int_{0}^{K} \frac{dx}{\sqrt{1-x^2}}
\exp\Big(-\frac{\beta^2}{1+x}\Big).
\end{eqnarray}

It should be noted that Eq.~\eqref{J_J0+int} can be also obtained by
means of the two-point probability density that for the correlated
Gaussian sequence $\beta_n$ with the correlator $K_\beta(r)$ is
defined by
\begin{eqnarray}\label{Gauss2PD}
&&\rho_B(\beta_n=\beta,\beta_{n+r}=\beta')\\
&=&\frac{1}{2\pi\sqrt{1-K_\beta^2(r)}}
\exp\left\{-\frac{{\beta}^2+{\beta'}^2-
2K_\beta(r)\beta\beta'}{2[1-K_\beta^2(r)]} \right\}.\nonumber
\end{eqnarray}

The last step we should take, is to calculate
$\mathcal{J}(0,\beta)$. It can be directly obtained from
Eq.~\eqref{J-def},
\begin{eqnarray}%\label{}
\mathcal{J}(0,\overline{\beta})=\left[\frac{2}{\pi}\int_{0}^\infty
dx \frac{\sin(\overline{\beta}x)}{x}
\exp(-x^2/2)\right]^2=\overline{\gamma}^2.
\end{eqnarray}
This result can be easily explained. Indeed, the condition
$K_\beta(r)=0$ implies that the correlations between $\beta_n$ and
$\beta_{n+r}$ disappear, hence, the correlations between $\gamma_n$
and $\gamma_{n+r}$ are absent as well.

As a result of these calculations, we finally get
\begin{eqnarray}%\label{}
\notag \mathcal{J}(K_\beta(r),\overline{\beta})
=\overline{\gamma}^2&+&\frac{2}{\pi}\int_{0}^{{K}_{\beta}(r)}
\frac{dx}{\sqrt{1-x^2}}
\exp\Big(-\frac{\overline{\beta}^2}{1+x}\Big)
\end{eqnarray}
This expression provides Eq.~\eqref{Corr}.

%%%%%%%%%%%%%%%%%%%%%%%%%%%%%%%%%%%%%%%%%%%%%%%%%%%%%%%%%%%%%%%%%%

%%%%%%%%%%%%%%%%%%%%%%%%%%%%%%%%%%%%%%%%%%%%%%%%%%%%%%%%%%%%%%%%%%

\end{document}